# Magnetic Resonance Force Microscopy with Overlapping Frequencies of Cantilever and Spin

Gennady P. Berman[1] and Vladimir I. Tsifrinovich[2]

[1]Theoretical Division, T-4, Los Alamos National Laboratory, Los Alamos, NM 87545, USA

[2]Department of Applied Physics, NYU Tandon School of Engineering, Brooklyn, NY 11201, USA

**Abstract**

We have studied theoretically magnetic resonance force microscopy (MRFM) with a high frequency nanomechanical cantilever when the cantilever frequency matches the resonant frequency of a single electron spin. Our estimations show that in this scenario the relative frequency shift of the cantilever can be much greater than the record MRFM frequency shift achieved in experiments with a single spin detection. Experimental realization of our proposal could open the way for fast detection of a single electron spin and even for detection of a single nuclear spin.

## 1. Introduction

In 2004 Rugar et al. demonstrated detection of an electron single spin below the surface of a nontransparent solid [1]. The authors of [1] used magnetic resonance force microscopy (MRFM) with oscillating cantilever-driven adiabatic reversals (OSCAR) technique. In this technique, oscillations of a low frequency micromechanical cantilever cause adiabatic reversals of a spin. The back magnetic force, produced by the spin on the cantilever, causes the frequency shift of the cantilever vibrations which can be detected with a high accuracy.

In the last ten years a significant progress was achieved in fabrication of high frequency nanomechanical cantilevers [2,3]. A similar progress was achieved in fabrication of magnetic nanoparticles which can be attached to the nanomechanical cantilever tip [4]. In this paper, we consider an opportunity to use a high frequency nanomechanical cantilever with a magnetic nanoparticle in MRFM. On our opinion, the most promising opportunity is associated with a possible matching the resonant spin frequency (the Larmor frequency) with the cantilever frequency. We claim that in this situation the cantilever frequency shift may significantly exceed the record shift achieved in a single spin measurement with the OSCAR technique.

## 2. The Hamiltonian of the CT-Spin System

We will consider a situation shown in Fig. 1: a uniform spherical ferromagnetic particle is attached to the tip of a vertical cantilever which can oscillate near the horizontal surface of a sample. A single non-paired electron spin of an atom or ion is located below the surface of the sample. The position of the center of the ferromagnetic particle will be referred as the position of the cantilever tip (CT) which can oscillate along the horizontal $x$–axis. The energy of the magnetic interaction between the ferromagnetic particle and the spin will be referred as the CT-spin interaction. With no CT-spin interaction the equilibrium position of CT is the origin. The spin is located at a point, $P$, with coordinates, $(0,0,-d)$, where $d$ is the equilibrium distance between the spin and CT with no CT-spin interaction.

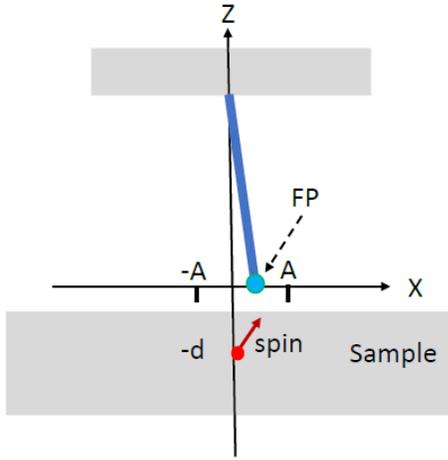

**Fig. 1**: The cantilever-spin system. Spin is located at a point: $(0,0,-d)$. FM is the ferromagnetic particle. "A" is the amplitude of CT oscillations.

First, we consider the magnetic dipole field, $\vec{B}(x_c)$, at the spin location produced by the particle, where $x_c$ is the coordinate of CT. It is clear that the displacement of CT by $x_c$ is equivalent to the displacement of the spin by $x = -x_c$. In the latter case we have,

$$B_j(x) = B_j(0) - G_{jx} x,$$
$$G_{jx} = -\left.\frac{\partial B_j}{\partial x}\right|_P, \quad j = x, y, z. \tag{1}$$

The classical energy of the CT-spin interaction can be written as,

$$U = -\vec{\mu}\cdot\vec{B} = -\sum_j \mu_j B_j(0) + \sum_j \mu_j G_{jx} x. \tag{2}$$

Here $\vec{\mu}$ is the spin magnetic moment. We will assume that the magnetic moment of the ferromagnetic particle $\vec{m}$ points in the negative $z$–direction: $\vec{m} = (0,0,-m)$. Then, in the equilibrium position with no CT-spin interaction only the $z$–component of the dipole magnetic field is not equal to zero at point $P$: $\vec{B}(0) = (0,0,-B_d)$, where $B_d$ is the magnitude of the dipole magnetic field at point, $P$ produced by the ferromagnetic particle located at the origin. Also, in the absence of an electric current the curl of the magnetic field near the spin location is zero. Using these two statements, we can re-write the expression for the CT-spin energy in more convenient form,

$$U = \mu_z B_d + \left(G_{xx}\mu_x + G_{xy}\mu_y + G_{xz}\mu_z\right)x. \tag{3}$$

The magnetic field produced by a uniformly magnetized spherical particle located at the origin is well known [5]:

$$\vec{B}(\vec{r}) = \frac{\mu_0}{4\pi r^3}\left\{\frac{3(\vec{m}\cdot\vec{r})\vec{r}}{r^2} - \vec{m}\right\}. \tag{4}$$

Here $\mu_0$ is the permeability of free space. The $x$–component of the magnetic field equals,

$$B_x = -\frac{3\mu_0 m}{4\pi r^5}zx. \tag{5}$$

From this equation, we find the partial derivatives in (1), and write the expression for the CT-spin energy in the form,

$$U = \mu_z B_d - G_{xx}\mu_x x_c,$$
$$G_{xx} = \frac{3\mu_0 m}{4\pi d^4}, \quad B_d = \frac{2\mu_0 m}{4\pi d^3}. \tag{6}$$

Note, that we changed $x$ to (-$x_c$). Next, we assume that an external magnetic field $\vec{B}_{ex}$ ($B_{ex} > B_d$) which points in the positive z-direction tunes the spin Larmor frequency $2\mu_B(B_{ex} - B_d)/\hbar$ to the CT frequency $\omega = k/M$, where $\mu_B$ is the Bohr magneton, $k$ and $M$ are the force constant and effective mass of CT. The energy of interaction between the spin and the external magnetic field is $-\mu_z B_{ex}$. The effective CT mass, $M$, equals to the sum of the effective mass of the CT with no ferromagnetic particle, $M_{np}$, and the mass of the particle, $M_p$ [3, 6]:

$$M = M_{np} + M_p. \tag{7}$$

Now, we can write the total quantum Hamiltonian of the CT-spin system [6]. We will express it in terms of the creation, $a^\dagger$, and the annihilation, $a$, operators:

$$H = \hbar\omega\left(a^\dagger a + \frac{1}{2}\right) + \frac{\hbar\omega}{2}\sigma_z + \hbar\lambda\sigma_x\left(a^\dagger + a\right),$$

$$a = \sqrt{\frac{M\omega}{2\hbar}}\left(x_c + \frac{i}{M\omega}p_c\right),$$

$$\lambda = \frac{\mu_B G_{xx}}{\sqrt{2\hbar\omega M}}.$$

(8)

Here $p_c$ is the CT momentum operator, $\sigma_z$ and $\sigma_x$ are the Pauli operators. The first term in the Hamiltonian (8) describes the energy of the CT oscillations, the second term is the spin Zeeman energy, the third one is the energy of the CT-spin interaction with the interaction constant, $\lambda$. The signs in the second and third terms of the Hamiltonian are positive due to the negative gyromagnetic ratio of an electron.

Hamiltonian (8) is the same as the Hamiltonian of the well-known Rabi model (see, for example, [7]). If the interaction constant, $\lambda$, is small compared to the frequency, $\omega$, the Rabi model can be reduced to the Jaynes-Cummings model with Hamiltonian,

$$H = \hbar\omega\left(a^\dagger a + \frac{1}{2}\right) + \frac{\hbar\omega}{2}\sigma_z + \frac{\hbar\lambda}{2}\left(a^\dagger \sigma_- + a\sigma_+\right),$$

$$\sigma_\pm = \sigma_x \pm i\sigma_y.$$

(9)

### 3. Estimation of the frequency shift

The energy levels of the Jaynes-Cummings model are well known [7]:

$$E_\pm(n) = \hbar\left\{\omega\left(n + \frac{1}{2}\right) \pm \frac{1}{2}\lambda\sqrt{n+1}\right\}.$$

(10)

Thus, the frequency shift of the CT oscillations can be estimated as,

$$\delta\omega(n) = \frac{1}{2}\lambda\sqrt{n+1} = \mu_B G_{xx}\sqrt{\frac{n+1}{8M\hbar\omega}}.$$

(11)

The mean value of $n$ can be expressed in terms of the quasiclassical amplitude, $A$, of the CT oscillations. Indeed, the mean CT energy is,

$$\langle E \rangle = \left( \langle n \rangle + \frac{1}{2} \right) \hbar \omega = \frac{1}{2} M (\omega A)^2. \tag{12}$$

Here angular brackets denote quantum mechanical average. Assuming that $\langle n \rangle \gg 1$, we obtain the expression for the relative frequency shift:

$$\frac{|\delta \omega|}{\omega} = \frac{\mu_B A}{4 \hbar \omega} \cdot G_{xx}. \tag{13}$$

This expression should be compared to the corresponding expression in the OSCAR MRFM, where the $z$ – component of the spin oscillates with the CT frequency, $\omega$, while the resonant Larmor frequency is much greater than $\omega$ [6]:

$$\frac{|\delta \omega|}{\omega} = \frac{2 \mu_B}{\pi k A} \cdot |G_{zx}| \tag{14}$$

The two expressions, (13) and (14), describe different dependences of parameters. In particular, the relative frequency shift (13) is proportional to the amplitude of oscillations, $A$, while the OSCAR shift (14) is inversely proportional to $A$.

Finally, we will estimate the value of the relative frequency shift (13). We note that the maximum magnetic field gradient, produced by a magnetic nanoparticle, can be much greater than the same gradient produced by a particle of a micrometer size. Indeed, if the radius of a spherical particle is $R$, the theoretical limit for the minimum CT-spin distance is $d = R$. The magnetic field gradient is proportional to the magnetic moment of the particle, $m$, which is proportional to $R^3$: $m \propto R^3$. From the other side, the gradient of the magnetic field is inversely proportional to $d^4 \propto R^4$. Thus, the maximum possible magnetic field gradient increases with decrease of $R$. Note, that in our estimations we are not going to take advantage from this opportunity. We will assume that a magnetic nanoparticle of radius 50 nm is attached to a cantilever tip. We will take the value of magnetization (per unit mass) reported in [4] for $Zn_xFe_{3-x}O_4$: $m/M_p = 200$ JT$^{-1}$kg$^{-1}$. Taking the density of iron oxide 5170 kg/m$^3$, we obtain the mass of the particle $M_p = 2.71 \cdot 10^{-18}$ kg, and its magnetic moment, $m = 5.41 \cdot 10^{-16}$ J/T. For the CT-spin distance, $d = 4R = 200$ nm, we obtain from (6):

$$G_{xx} = \frac{3 \mu_0 m}{4 \pi d^4} = 1.014 \cdot 10^5 \text{ T/m}. \tag{15}$$

This value is close to the corresponding typical values in the OSCAR experiments. As an example, in ref. [8],

$$G_{zx} = 4.3 \cdot 10^5 \text{ T/m}. \tag{16}$$

We will consider a $600x400x100$ nm silicon carbide cantilever reported in [2], with the frequency with no particle, $\omega_{np}/2\pi = 127$ MHz, and force constant, $k = 32.1$ N/m. From this data, we can find the effective mass of the CT with no ferromagnetic particle,

$$M_{np} = \frac{k}{\omega_{np}^2} = 5.04 \cdot 10^{-17} \text{ kg}. \tag{17}$$

The effective mass of the CT with a ferromagnetic particle is,

$$M = M_{np} + M_p = 5.31 \cdot 10^{-17} \text{ kg}. \tag{18}$$

Correspondingly, the frequency of the CT is,

$$\frac{\omega}{2\pi} = \frac{1}{2\pi}\sqrt{\frac{k}{M}} = 124 \text{ MHz}. \tag{19}$$

Finally, taking the amplitude of oscillations, $A = 10$ nm, which is similar to that used in the OSCAR experiments (e.g., $A = 16$ nm, in [1], and $A = 10$ nm, in [8]), we compute the relative frequency shift (13):

$$\frac{|\delta\omega|}{\omega} = \frac{\mu_B A}{4\hbar\omega} \cdot G_{xx} = 2.86 \cdot 10^{-2}. \tag{20}$$

The relative frequency shift in the OSCAR single spin detection [1] was $7.64x10^{-7}$, more than 4 orders of magnitude smaller than the value estimated in (20). Thus, we can expect that matching the CT and spin resonant frequencies can significantly simplify the detection of a single electron spin, and even open the way for detection of a single nuclear spin.

4. **Conclusion**

We suggested to use recent advance in fabrication of high frequency nanomechanical cantilevers and magnetic nanoparticles in order to match the Larmor spin frequency to the frequency of the cantilever. We have found a situation when the Hamiltonian of the CT-spin system is exactly the same as the Hamiltonian of the Rabi model which can be reduced to the Jaynes-Cummings model. Our estimations show that when the spin Larmor frequency matches the CT frequency the frequency shift of the CT oscillations caused by a single spin is more than four orders of magnitude greater than the similar shift demonstrated with the OSCAR MRFM technique. The experimental realization of our proposal would significantly simplify detection of a single electron spin and open the way for detection of a single nuclear spin.

## Acknowledgements

The work by G.P.B. was done at Los Alamos National Laboratory managed by Triad National Security, LLC, for the National Nuclear Security Administration of the U.S. Department of Energy under Contract No. 89233218CNA000001.